# The Open Cloud Testbed: A Wide Area Testbed for Cloud Computing Utilizing High Performance Network Services

Robert Grossman[1,2], Yunhong Gu[1], Michal Sabala[1],
Collin Bennett[2], Jonathan Seidman[2] and Joe Mambratti[3]

[1] National Center for Data Mining, University of Illinois at Chicago

[2] Open Data Group

[3] International Center for Advanced Internet Research, Northwestern University

## ABSTRACT

Recently, a number of cloud platforms and services have been developed for data intensive computing, including Hadoop, Sector, CloudStore (formerly KFS), HBase, and Thrift. In order to benchmark the performance of these systems, to investigate their interoperability, and to experiment with new services based on flexible compute node and network provisioning capabilities, we have designed and implemented a large scale testbed called the Open Cloud Testbed (OCT). Currently the OCT has 120 nodes in four data centers: Baltimore, Chicago (two locations), and San Diego. In contrast to other cloud testbeds, which are in small geographic areas and which are based on commodity Internet services, the OCT is a wide area testbed and the four data centers are connected with a high performance 10Gb/s network, based on a foundation of dedicated lightpaths. This testbed can address the requirements of extremely large data streams that challenge other types of distributed infrastructure. We have also developed several utilities to support the development of cloud computing systems and services, including novel node and network provisioning services, a monitoring system, and a RPC system. In this paper, we describe the OCT architecture and monitoring system. We also describe some benchmarks that we developed and some interoperability studies we performed using these benchmarks.

## 1. INTRODUCTION

Cloud computing has become quite popular during the last few years. There are several reasons for this. First, the popular cloud computing services are very easy to use. Users can request computing resources on demand from cloud service providers. Also, most users find the Hadoop Distributed File System (HDFS) and the Hadoop implementation of MapReduce [8] very easy to use compared to traditional high performance computing programming frameworks, such as MPI. Second, it is relatively straightforward to set up a basic cloud computing facility. All that is required are one or more racks containing commodity servers. Third, the payment model provides advantages to users with variable resource requirements. For example, it allows for quick implementation of processing resources without an upfront capital investment.

Cloud computing systems are commonly divided into three types (Figure 1): 1) on-demand virtual machine images, such as Amazon's EC2 [5] or Eucalyptus [4], which are known as Infrastructure as a Service (IaaS); 2) storage, data processing, and compute services such as those provided by S3 [5], Hadoop [2], CloudStore [3], and Sector/Sphere [1] that can be used to build applications, which are known as Platform as a Service (PaaS); and 3) software applications delivered as a service, such as Google apps, which are known as Software as a Service (SaaS).

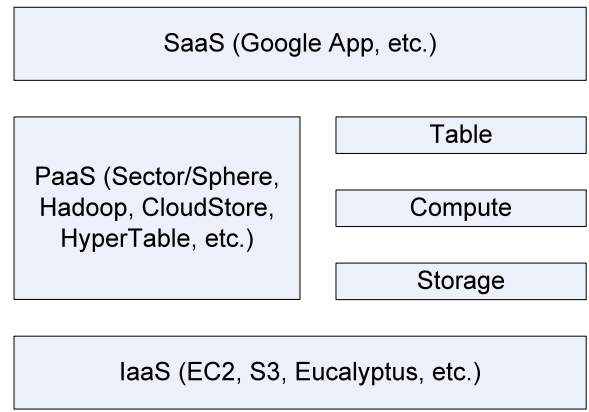

**Figure 1. Cloud Computing Stack.**

With all these different cloud systems and services, there is clear need for a testbed for benchmarking different systems and testing their interoperability. The Open Cloud Testbed (OCT) was designed to benchmark cloud systems, to investigate their interoperability, and to experiment with implementations on novel infrastructure, such as large scale high performance optical networks. Also, networks are integrated as "first class" controllable, adjustable resources not merely as external resources.

In the OCT, a variety of cloud systems and services are installed and available for research, including Eucalyptus [4], Hadoop [2], CloudStore (KosmosFS) [3], Sector/Sphere [1], and Thrift [6]. Supporting these systems and services makes it easier to perform interoperability and benchmarking studies.

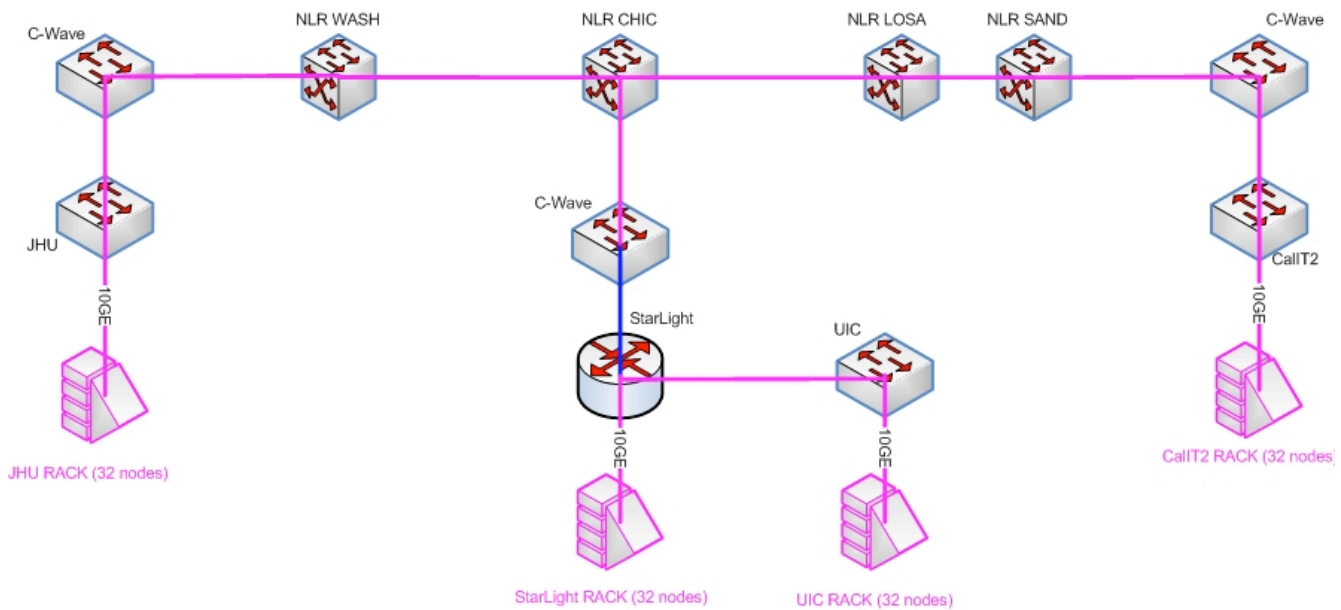

**Figure 2: The Open Cloud Testbed System Diagram**

In addition, we have also developed network libraries, monitoring systems, and benchmark suites to support the development and experimental studies of cloud computing stacks.

In this paper, we describe the OCT architecture and monitoring software. We also describe some benchmarks that we have developed and preliminary results of experimental studies that we have conducted using these benchmarks.

## 2. THE OPEN CLOUD TESTBED
### 2.1 Concepts and Objectives
The OCT architecture envisions the emergence of powerful large scale distributed applications supported by services and processes based on highly distributed, integrated facilities and infrastructure. In this view, the infrastructure is an extremely flexible programmable platform that allows new functions and capabilities to easily be created and implemented.

The OCT represents a departure from existing clouds is several ways. First, as its name implies, it is based on a concept of interoperability and openness. Second, most clouds today use standard protocols to support their services and infrastructure. The OCT architecture incorporates high performance services, protocols, and infrastructure at all levels. Instead of using the commodity Internet, it uses a national high performance 10 Gb/s network based on extremely fast transport protocols supported by dedicated light paths. Although such capabilities are fairly rare today, this approach is being used to model future distributed infrastructure, which will provide much more capacity and capabilities than current systems. For example, as commonly implemented, clouds do not support large data streams well. In contrast, OCT is being designed not only to manage millions of small streams and small amounts of information but also extremely large data sets and very high volume data streams.

The objectives of the OCT initiative extend beyond creative novel high performance capabilities. Another important research objective is to develop standards and frameworks for interoperating between different cloud software. Currently, we have tested Eucalyptus, CloudStore (formerly KosmosFS), Hadoop, Sector/Sphere, and Thrift. In particular, we developed an interface so that Hadoop can use Sector as its storage system.

### 2.2 The OCT Infrastructure
As illustrated in Figure 2, currently there are 4 racks of servers in the OCT, located in 4 data centers at Johns Hopkins University (Baltimore), StarLight (Chicago), the University of Illinois (Chicago), and the University of California (San Diego). Each rack has 32 nodes. Each node has dual dual-core AMD 2.4GHz CPU, 12GB memory, 1TB single SATA disk, and dual 1GE NICs. Two Cisco 3750E switches connect the 32 nodes, which then connects to the outside by a 10Gb/s uplink.

In contrast to other cloud testbeds, the OCT utilizes wide area high performance networks, not the familiar commodity Internet. There is a 10Gb/s network that connects the various data centers. This network is provided by the CiscoWave national testbed infrastructure, which spans the US east, west, north and south. The majority of experimental studies we perform on the OCT extend over all four geographically distributed racks. In contrast, most experimental studies using cloud computing take place within a single data center.

We are currently installing two more racks (60 nodes) and expect to install two additional racks by the end of 2009. By then the OCT will have about 250 nodes and 1000 cores. We are also extending the 10Gb/s network to MIT Lincoln Lab (Cambridge) and the Pittsburgh Supercompter Center/Carnegie Mellon University (Pittsburgh).

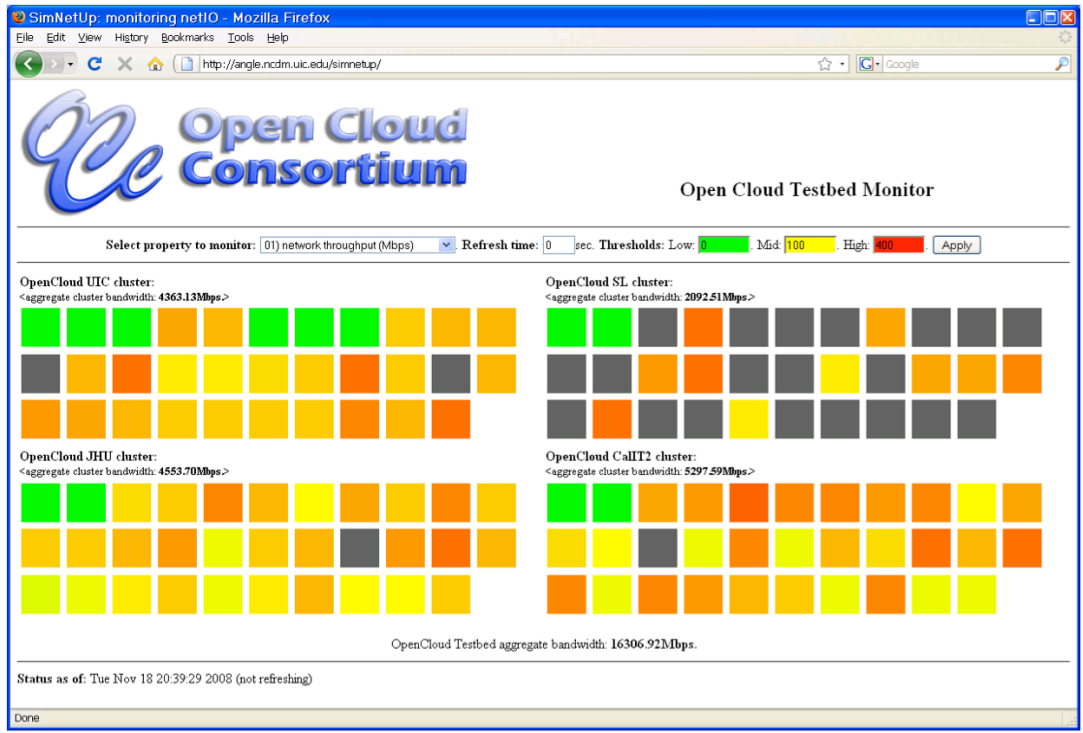

**Figure 3. The Open Cloud Testbed monitoring and visualization system.**

## 3. MONITORING AND VISUALIZATION

Developing, debugging, and studying the performance of distributed systems is quite complex. Management costs increase significantly as the number of sites increases. To counter this, we have developed a simple but effective monitoring and real time visualization system to track the behavior of every node in the OCT [11]. The OCT monitoring system records the resource utilization (including CPU, memory, disk, NIC, etc.) on each node. A web-based visualization (Figure 3) allows users to easily and directly monitor the entire system.

In addition, the monitoring system also helps with benchmarking and debugging cloud software. The visualization effectively indicates the real time status of the testbed – for example, if it is fully utilized and if the load is balanced across the testbed. This capability has proved invaluable when debugging software on the testbed.

A snapshot of the web-based visualization can be found in Figure 3. Each block represents a server node, and each group of blocks represent a cluster. The color of each block represents the usage of a particular resource, in this case, network IO throughput. Color on the green/light side means the machine is idle; color on the red/dark side means the machine is busy.

In the Sector/Sphere software, there is also built-in a monitoring system that is used to improve load balancing and to remove nodes and/or network segments that exhibit poor performance. While the OCT monitoring system records information for the node as a whole, the built-in monitoring system of Sector records the resource usage of the Sector process on each node and Sector's network utilization.

Sector assumes that the underlying network has a hierarchical topology, such as the one used by OCT. Based on this topology, Sector computes the aggregate network throughput on each link, in addition to each node. This process helps to identify a malfunctioning link or node and in this way Sector can remove underperforming resources from the system.

The OCT architecture is based on an assumption that services are based on flexible, not static, foundation resources. The architecture provides for the continuous monitoring of conditions, and the identification of changes, including at the level of network resources. The OCT leverages and builds on recent research and development trends in architecture and technologies that provide for dynamically provisioned network resources [13]. In contrast, most architectures today are based on static network resources.

## 4. GMP: A MESSAGING PROTOCOL

We have also designed and developed a new open source high performance messaging protocol called GMP (Group Messaging Protocol) for Sector and other distributed systems. GMP allows an application to send a message quickly and reliably to another node. High performance messaging is essential for rapid reconfigurations of core resources under changing conditions. The protocol is suitable for delivering small control messages in distributed systems.

GMP is a connection-less protocol, which uses a single UDP port and which can send messages to any GMP instances or receive messages from other GMP instances. Because there is no connection setup required, GMP is much faster than TCP, which requires a connection to be set up between the communicating nodes. GMP, which is built on top of UDP, does not maintain virtual connections, but instead maintains a list of states for each peer addresses to which it sends messages or from which it receives messages.

Every GMP message contains a session ID and a sequence number. Upon receiving a message, GMP sends back an acknowledgment; if no acknowledgment is received, the message will be sent again. A mechanism like this is required since UDP is an unreliable protocol. The sequence number is used to make sure that no duplicated message will be delivered. The session ID is used to differentiate messages from the same address but different processes (e.g., if one process is restarted it will use a different session ID). If the message size is greater than a single UDP packet can hold, GMP will set up a UDT [12] connection to deliver the large message. However, we expect such a situation to be rare for GMP.

In Sector, we also developed a light-weight high performance RPC mechanism on top of GMP. The RPC library simply sends out a request in a GMP message and then it waits for the response to come back.

## 5. BENCHMARKS

We have developed several benchmarks for evaluating applications and services for large data clouds. In particular, we have developed a benchmark called MalStone [14].

MalStone is a stylized analytic computation that requires the analysis of log files containing events about entities visiting sites of the following form:

| Event ID | Timestamp | Site ID | Compromise Flag | Entity ID

The assumption is that some of the entities that visit certain sites become compromised. In order to find out the compromised sites that compromise entities, the benchmark application requires that a ratio be computed *for each site* that for a specified time window measures the percent of entities that become compromised at any time in the window.

There are several variants of MalStone. MalStone-A computes the overall ratio per site. Malstone-B computes a series of windows-based ratio per site. MalStone is commonly used with 10 billion, 100 billion or 1 trillion 100-byte records (so that there is 1 TB, 10 TB and 100 TB of data in total).

This type of computation requires only a few lines of code if the data is on a single machine (and can be done easily in a database). On the other hand, if the data is distributed over the nodes in a cloud, then we have found this type of computation turns out to be a useful benchmark for comparing different storage and compute services.

An example of a situation that might result in these types of log files is what are sometimes termed drive-by exploits [10]. These incidents result when users visit web sites containing code that can infect and compromise vulnerable systems. Not all visitors become infected but some do.

We have also developed a data generator for the MalStone benchmark called MalGen [14].

## 6. EXPERIMENTAL STUDIES

In this section, we describe several experimental studies we have conducted on the OCT.

In the first series of experiments, we used MalGen to generate 500 million 100-byte records on 20 nodes (for a total of 10 billion records or 1 TB of data) in the OCT and compared MalStone's performance using: 1) the Hadoop Distributed File System (HDFS) with Hadoop's implementation of MapReduce; 2) HDFS with Streams and MalStone coded in Python; and, 3) the Sector Distributed File System (SDFS) and MalStone coded using Sphere's User Defined Functions (UDF). The results are below (Table 1):

|  | Malstone-A | MalStone-B |
|---|---|---|
| Hadoop MapReduce | 454m 13s | 840m 50s |
| Hadoop Streams with Python | 87m 29s | 142m 32s |
| Sector/Sphere | 33m 40s | 43m 44s |

Table 1. Hadoop version 0.18.3 and Sector/Sphere version 1.20 were used for these tests. Times are expressed in minutes (m) and seconds (s).

The Sector/Sphere benefits significantly from mechanisms that optimize data movement (bandwidth load balancing and UDT data transfer) and it performs much faster than Hadoop.

|  | 28 Local Nodes | 7 * 4 Distributed Nodes | Wide Area Penalty |
|---|---|---|---|
| Hadoop (3 replicas) | 8650 | 11600 | 34% |
| Hadoop (1 replica) | 7300 | 9600 | 31% |
| Sector | 4200 | 4400 | 4.7% |

Table 2. This table compares the performance of Hadoop and Sector for a computation performed in one location using 28 nodes and 4 locations using 7 nodes each.

In the second series of experiments, we used MalGen to 15 billion on 28 nodes in one location and compared these results to the same computation performed when the nodes were distributed over four locations in the testbed. See Table 2. The experiment shows the impact of wide area networks on the performance of such applications. The performance penalty on Hadoop is 31~34%, while Sector suffers a 4.7% performance drop.

There are two major reasons for the better performance of Sector over wide area networks. First, Sector employs a load balancing mechanism to smoothly distribute the network traffic within the system. Second, Sector uses UDT [12] for data transfer. UDT is a high performance protocol that performs significantly better than TCP over wide area networks. The limitations of TCP are well documented [13].

## 7. RELATED WORK

There are several other cloud computing testbeds and platforms that serve similar or related goals as that of the OCT. In this section, we describe four of these. There are also a variety of testbeds available for other types of research, such as high performance computing and high performance computing applications (e.g. the Teragrid), network research (e.g. PlanetLab), and network security research (e.g. DETER), that we do not describe.

The most closely related one is the Open Cirrus Testbed [9]. Open Cirrus consists of 6 sites with various numbers of nodes, ranging between 128 and 480 per site. While Open Cirrus contains many more nodes than OCT, it is not designed to support computations that span more than one data center. That is, the Open Cirrus Testbed is designed for systems that run within a single data center.

Another cloud computing testbed is the Google-IBM-NSF Cluster Exploratory or CluE Resource. The Google-IBM-NSF CluE resource appears to be a testbed for cloud computing applications in the sense that Hadoop applications can be run on the testbed but that the testbed does not support systems research and experiments involving cloud middleware and cloud services, as is possible with the OCT and the Open Cirrus Testbed.

The Eucalyptus Public Cloud [15] is a testbed for Eucalyptus applications. Currently, users are limited to no more than 4 virtual machines and experimental studies that require 6 hours or less.

Amazon's EC2 platform [5] provides an economical alternative for certain types of cloud computing research. However, in EC2 users cannot control data locality and network configurations. This makes foundational research on cloud systems and services difficult. In addition, while EC2 is inexpensive for temporary use, it can be quite expensive as the scale of the experiments and the duration increase.

## 8. CONCLUSION

Cloud computing has proven to be an important platform for many types of applications, including applications oriented to consumers, applications supporting the enterprise, and applications for large scale science. This trend is expected to continue for the foreseeable future. However, in many ways research in cloud computing is still immature. There is limited understanding of basic issues, such as exploiting data locality, load balancing, and identifying nodes and collections of nodes that are damaging the overall performance of an application. Also, there are not yet commonly accepted standards to support interoperability. Nor are there well established cloud computing benchmarks.

The OCT initiative was established in order to advance cloud computing research, to enable experimental studies involving different cloud computing architectures, to provide a platform for cloud interoperability studies, and to encourage cloud computing benchmarks.

The OCT is the only cloud computing testbed that utilizes wide area, high performance networks. It also supports experimental studies using non-TCP protocols, such as UDT, and with dynamic services based on flexible compute node and network provisioning capabilities.

The OCT monitoring and visualization system has proved to be critical to completing a number of experimental studies and to identifying directions for future research. For example, it was through this system that the sometimes dramatic impact on an application of just one or two nodes with slightly inferior performance was first noted. In future work, we plan on improving our understanding of this issue. We are currently completing several interoperability studies technical reports are under preparation describing them.

As mentioned above, the OCT is managed by the Open Cloud Consortium (OCC). Information about joining the Open Cloud Consortium and participating in cloud computing research studies, interoperability studies, and benchmarking studies is available from the OCC web site (www.opencloudconsortium.org).

## 9. Acknowledgements

The Open Cloud Testbed is managed and operated by the Open Cloud Consortium. Partial support for this work was provided by the Open Cloud Consortium and its members and by the National Science Foundation (NSF) through NSF grants OCI-0430781, CNS-0420847, and ACI-0325013.